\documentclass[aps,showpacs,twocolumn,floatfix,prc,superscriptaddress]{revtex4}
\usepackage{epsfig,graphics,color}
\usepackage{graphicx}% Include figure files
\usepackage{dcolumn}% Align table columns on decimal point
\usepackage{bm}% bold math
\usepackage{amsmath}
\usepackage{eucal,eufrak}

\newcommand \tie {{\it i.e.}}
\newcommand \ie {{\it i.e.} }
\newcommand \f {\not\!}

\newcommand \kd  {\delta}
\newcommand \ra  {\rightarrow}

\newcommand \vk {\vec{k}}
\newcommand \vl {\vec{l}}

\newcommand \g {\gamma}

\newcommand \x {\cdot}
\newcommand \hf {\frac{1}{2}}
\newcommand \A {\alpha}
\newcommand \B {\beta}

\newcommand \lc {\langle}
\newcommand \rc {\rangle}
\newcommand \prt {\partial}

\newcommand \bvec{\left( \begin{array}{c} }
\newcommand \evec{\end{array} \right)}
\newcommand \tr {\mbox{{\bf Tr}}}
  
\newcommand \bea{\begin{eqnarray} }
\newcommand \eea{\end{eqnarray} }
\newcommand \nn {\nonumber}
\newcommand \be {\begin{equation}}
\newcommand \ee {\end{equation}}

\newcommand \mbx {\mbox{}}

\newcommand \psibar {\bar{\psi}}
\newcommand \ata {& \times &}

\begin{document}

\title{Elastic energy loss and longitudinal straggling of a hard jet}

\author{A. Majumder}
\affiliation{Department of Physics, Duke University, Durham, NC 27708, USA}
\affiliation{Department of Physics, Ohio State University, Columbus, OH 43210, USA}

% \author{B. M\"uller}
% \affiliation{Department of Physics, Duke University, Durham, NC 27708, USA}

\date{ \today}

\begin{abstract} 
The elastic energy loss encountered by  jets produced in deep-inelastic scattering (DIS) off a large 
nucleus is studied in the collinear limit. In close analogy to the case of (non-radiative) transverse momentum broadening, which is dependent on the medium transport coefficient $\hat{q}$, a class of medium enhanced higher twist operators which contribute to the non-radiative loss of the forward light-cone momentum of the jet ($q^-$) are identified and the leading correction in the limit of asymptotically high $q^-$ is isolated. Based on these operator products, a new transport coefficient $\hat{e}$ is motivated which quantifies the energy loss per unit length encountered by the hard jet. 
These operator products are then computed, explicitly, in the case of a similar hard jet traversing a 
deconfined quark-gluon-plasma (QGP) in the hard-thermal-loop (HTL) approximation. This is followed by an 
evaluation of sub-leading contributions which are suppressed by the light-cone momentum $q^-$, 
which yields the longitudinal ``straggling'' i.e., a slight change in light cone momentum due to the Brownian propagation 
through a medium with a fluctuating color field. 
\end{abstract}

\pacs{12.38.Mh, 11.10.Wx, 25.75.Dw}

\maketitle

%%%%%%%%%%%%%%%%%%%%%%%%%%%%%%%%%%%%%%%%%%%%%%%%%%%%%%%%%%
%%%%%%%%%%%%%%%%%%%%%%%%%%%%%%%%%%%%%%%%%%%%%%%%%%%%%%%%%%
%%%%%%%%%%%%%%%%%%%%%%%%%%%%%%%%%%%%%%%%%%%%%%%%%%%%%%%%%%
%%%%%%%%%%%%%%%%%%%%%%%%%%%%%%%%%%%%%%%%%%%%%%%%%%%%%%%%%%
%%%%%%%%%%%%%%%%%%%%%%%%%%%%%%%%%%%%%%%%%%%%%%%%%%%%%%%%%%

%  \section{introduction}

%%%%%%%%%%%%%%%%%%%%%%%%%%%%%%%%%%%%%%%%%%%%%%%%%%%%%%%%%%
%%%%%%%%%%%%%%%%%%%%%%%%%%%%%%%%%%%%%%%%%%%%%%%%%%%%%%%%%%
%%%%%%%%%%%%%%%%%%%%%%%%%%%%%%%%%%%%%%%%%%%%%%%%%%%%%%%%%%
%%%%%%%%%%%%%%%%%%%%%%%%%%%%%%%%%%%%%%%%%%%%%%%%%%%%%%%%%%

The modification of hard jets as they propagate through dense matter is now a 
rather sophisticated enterprise~\cite{quenching,AMY,HT,GLV,ASW}, encompassing the 
study of jets propagating through cold nuclear matter in Deep-Inelastic scattering off large nuclei~\cite{Airapetian:2000ks}
as well as jets in hot deconfined matter produced in the collision of heavy-ions at high energy~\cite{white_papers}.
Unlike most approaches to jet modification, the Higher-Twist expansion approach~\cite{HT} attempts to 
decompose the measured modification into a part computable using 
perturbative QCD (pQCD) and a set of non-perturbative ``jet-transport-coefficients'' which are then 
used to quantify the 
properties of the dense matter. Since very few assumptions about the medium have been made, this formalism 
is equally applicable to both confined and deconfined matter. 
As a result, such a study has universal applications to a variety of heavy-ion experiments at the 
Relativistic Heavy-Ion Collider (RHIC) and the Large Hadron collider (LHC) and to DIS experiments at 
the Continuous Electron Beam Accelerator Facility (CEBAF) and the future Electron-Ion Collider (EIC).
To date, most jet modification calculations, in the higher twist scheme, 
have been limited to radiative energy loss and transverse broadening of the 
hard jet~\cite{Majumder:2007iu}. Both these signals of jet modification are dependent on the transverse transport 
parameter $\hat{q}$: which encodes the mean transverse momentum squared picked up by a 
hard jet per unit length of medium traversed~\cite{quenching}.

In this Note, the longitudinal momentum loss of the jet due to elastic exchange with the 
medium will be explored. We begin by first defining and clarifying the notion of elastic 
energy loss. It should be pointed out, that this is by far not the first attempt to explore 
this avenue of jet modification (see for instance Refs.~\cite{Thomas:1991ea,Djordjevic:2006tw,Wang:2006qr,Qin:2007rn}).
However, most of these computations assume a specific model of the medium: that of a thermalized 
plasma of quarks and gluons both within the HTL limit (as in Ref.~\cite{Thomas:1991ea,Djordjevic:2006tw,Qin:2007rn}) 
and without invoking such limits~\cite{Wang:2006qr}. Our formulation, in this manuscript, assumes no 
particular model of the medium and is thus related to that of Ref.~\cite{Wang:2006qr}. However, our results 
will differ from those of Ref.~\cite{Wang:2006qr} in that we identify a process which does \emph{not} interfere 
with radiative energy loss and depends on a new transport coefficient. In so doing, the manuscript 
strives to attain a non-ambiguous definition of elastic energy loss within the various scenarios that may be 
encountered in current experiments on DIS and heavy-ion collisions.  

It is no surprise that the magnitude of elastic energy loss (and even radiative energy loss) 
is a frame dependent statement. While in heavy-ion collisions, the natural frame of choice is the center of 
mass frame of the two colliding heavy-ions, there is more than one obvious choice in DIS on a large nucleus.
As a result, in this Note, the focus will lie on the approximately boost invariant quantity of fractional light-cone momentum loss, 
\bea
z = \frac{ E_{lost} + p^3_{lost} }{E +p^3},
\eea
where $E_{lost}$ and $p^3_{lost}$ refer to the energy and $z$-component of the momentum lost by a 
hard parton with energy and $z$-component of the momentum given by $E$ and $p^3$. The qualifier, approximately 
boost invariant, is used as the above quantity is only invariant for the case of boosts restricted solely to the $3$-direction, 
or the direction of the chosen component of momentum. Such a construction is somewhat alien to treatments of elastic energy 
loss in classical electrodynamics (See Chapter 13 of Ref.~\cite{jackson}), where the lost momentum is usually transverse 
to the direction of motion of the fast moving charge.

Imagine that a hard quark is produced in the DIS of an electron on a large nucleus. 
Such an analysis is carried out with maximal 
ease in the Breit frame, where both the virtual photon and the large nucleus approach each other 
at a large momentum~\cite{Majumder:2007hx}. In the interest of simplicity, we imagine that the 
nucleus moves in the positive $z$-direction and has a large  momentum $Ap^+$, where $A$ is the 
atomic number of the nucleus. A hard quark with light cone momentum $x_B p^+$ is struck by 
the virtual photon which moves in the negative $z$-direction with momentum 
\bea 
q_\g = \left[  -x_B p^+ , q^- , 0, 0  \right],
\eea
where, the Bjorken variable has the usual value $x_B = Q^2/(2p^+ q^-)$ in terms of the virtuality $Q^2$ of the 
virtual photon. In this frame, the struck quark has its four-momentum changed to $l = x_B p + q \simeq [0,q^-,0,0]$, may be 
thought of as close to on-shell and moving in a direction opposite to that of the large nucleus. The differential hadronic tensor 
for the production of an on-shell  
hard quark with a momentum $l$  in the interaction of a virtual photon on a large nucleus $A$ is given as
\bea
\frac{ d^3 W_0^{\mu \nu}  }{d^2 l_\perp d l^- } =  W^{\mu \nu}_0 \kd^2(\vl_\perp) \kd(l^- - q^- ),
\eea
where, $W^{\mu \nu}_0$ represents the inclusive hadronic tensor at leading order and leading twist, i.e.,
\bea 
W_0^{\mu \nu} = -C_p^A 2\pi g_\perp^{\mu \nu} \sum_q Q_q^2 f_q(x_B) , 
\eea
where, we have used the light-cone notation $- g_\perp^{\mu \nu} = g^{\mu +} g^{\nu -} + g^{\nu +} g^{\mu -} - g^{\mu \nu}$.
The coefficient $C_p^A$ simply counts the number of nucleons in the nucleus that the jet may scatter off and is equal to $A$.

In the remainder, we focus on the multiple soft reinteractions of this quark with the soft glue field within the various 
nucleons. The weakly interacting picture of nucleons in the large nucleus (which is approximately valid at very high energies) 
is now imposed (see Ref.~\cite{Majumder:2007hx} for details). Interactions of the quark with the various nucleons may 
henceforth be considered to be independent and thus uncorrelated. This step assumes a factorization of the production 
process of the hard quark and its later soft scattering off the glue field. As a result, a reader, more interested in the 
interactions of a quark with deconfined matter, may replace the individual nucleons 
with the prevalent degrees of freedom of the matter under study~\cite{Shuryak:2003ty,Koch:2005vg}.

Consider the soft rescattering of the hard quark with one such 
nucleon (or degree of freedom). 
In this limit of factorized interactions of the hard quark with the various nucleons, the effect of multiple 
interactions on the propagation of the hard quark may be iterated from the effect of a single interaction. 
This is not true in general and arguments will be forwarded which justify the use of this scheme in this 
particular case. By the effect on the hard quark, we specifically mean the effect on its three dimensional 
distribution in momentum space i.e., in terms of $l^-,\vl_\perp$. 
The aim is to identify the form of the distribution  $\phi_L(l^-,\vl_\perp)$, after the parton 
has traversed a certain length $L$ given an initial distribution, 
\bea
\phi_0(l^-,\vl_\perp) = \kd^2(\vl_\perp) \kd ( l^- - q^- ).
\eea

The hadronic tensor with two gluon scatterings in the final state, after a few simplifications,  may be expressed as 
%\begin{widetext}
%
\bea
W^{\mu \nu} \!\!\!\!&=& \!\!\!(-g_\perp^{\mu \nu }) C_{p,p_2}^A  \!\!\int d y_0^- e^{-ix_B p^+ y_0^-}  
\lc p | \psibar(y_0^-) \frac{\g^+}{2} \psi(0) | p \rc \nn \\
\ata \!\!g^2 \int \frac{d l^- d^2 l_\perp}{ (2\pi)^3 } d Y^- d y^- dy^+ d^2 y_\perp \frac{dk^- d^2 k_\perp}{(2\pi)^3} 
\nn \\
\ata \!\!\frac{(2\pi)^3 \kd(l^- - q^ - - k^- ) \kd^2 (\vl_\perp - \vk_\perp)}{2(q^- + k^-)}
\frac{\tr [t^a t^b]}{N_c} \nn \\ 
\ata \!\!\frac{\tr}{4} \left[ \g^+ \g^\A \left\{ (\f q^- \!+\f k^- ) + 
 \frac{ \g^- k_\perp^2}{2 (q^- \!+k^-)} - \f k_\perp  \right\} \right.  \nn \\
\ata \left. \g^\B  \right] \exp{ \left[ - i \frac{k_\perp^2}{2q^-} (y^-) + i y_\perp \x k_\perp -i y^+ k^- \right] } \nn \\
\ata \!\!\lc p_2 | A^a_\A (Y^- + y)    A^b_\B (Y^-)   | p_2 \rc. \label{W_mu_nu}
\eea
%
%\end{widetext}
%
In the limit of large energy, $q^- \ra \infty$, the distribution of $k^-$ which is obtained by Fourier transforming the 
$y^+$ dependence of the product $A^{\A} (Y^- + y) A^{\B} (Y^-) $ will be dominated by values of $k^- \ll q^-$. 
Under these limits, $k^-$ may be dropped from the momentum dependent part of the integrand [third and fourth line of 
Eq.~\eqref{W_mu_nu}] and 
the integral over $k^-$ may be performed using the phase factor $\exp{(-ik^- y^+)}$
which constrains the entire process to the negative lightcone. The resulting expression is then identical to the well known 
expression obtained for the standard treatment of transverse broadening with scattering off a single gluon in the 
final state~\cite{Guo:1998rd}. The coefficient $C^A_{p,p_2}$, where $|p\rc$ represents the nucleon with the struck 
quark and $| p_2 \rc$ the nucleon which contains the soft gluons, accounts for $A$ times the weak correlation between the 
two nucleons involved. This coefficient is derived in Ref.~\cite{Majumder:2008jy} and simplifications discussed in 
Ref.~\cite{Majumder:2007hx}.

In the evaluation of transverse broadening in Ref.~\cite{Majumder:2007hx}, the two 
dimensional delta function $\kd^2 (\vl_\perp - \vk_\perp)$ was expanded in 
a Taylor series in $k_\perp$ and the coefficients of each term of the expansion became the gluon matrix elements which 
appear in a product with the delta function. In the present case of non-radiative energy loss, we will follow a similar methodology and  
expand the delta function $\kd (l^- - q^- - k^-)$ as a series in $k^-$ and focus on the first derivative or the linear change 
in the distribution of $l^-$.  Hence, the delta function may now be reexpressed as 
\bea
\kd (l^- - q^- - k^-) &=& \kd(l^- - q^-) - \frac{ \prt \kd(l^- -q^-)  }{ \prt l^- } (k^-) \nn \\ 
&+&\frac{1}{2} \frac{\prt^2}{ \prt {l^-}^2} \kd (l^- - q^-) [k^-]^2  + \ldots \label{delta_func}
\eea

We use the equation above as a substitution for the $\kd$-function in Eq.~\eqref{W_mu_nu} and identify the 
coefficient ($C_1$) of the first derivative of the $\kd$-function as the magnitude of the collisional energy loss. This  may be schematically decomposed as,
\bea
C_1 = \mathcal{K} (k) \exp (k \x y) \lc  \mathcal{Y} (y)  \rc ,
\eea
\tie, a convolution of a purely momentum 
dependent part $\mathcal{K}$ with the expectation of a purely position dependent 
piece $\mathcal{Y}$ through a phase factor involving both.

In the limit of $|k^- | \ll q^-$  and small coupling constant $g$, the momentum dependent piece $\mathcal{K}$ may be expanded in a series in $k^-/q^-$, 
% and the spatial piece $\mathcal{Y}$ may be expanded in derivatives of $y^+$, 
\tie,
\bea
\mathcal{K} &\simeq& -k^- \left[ q^- g^{\A +} g^{\B +} + \hf \left(1 - \frac{k^-}{q^-} \right) \right. \nn \\
\ata \left. \left\{ g^{\A + } k_\perp^\B + g^{\B + } k_\perp^\A \right\}  
 -  \frac{k_\perp^2 }{4 q^-} \left(1 - \frac{k^-}{q^-} \right) g^{\A \B}_\perp \right]. \label{k_expand} 
\eea
The piece which depends on the coordinates may be simply expressed as $\mathcal{Y}\simeq{A^a}^{\A}(y) {A^b}^{\B} (0)$.
%where, the standard notation of $A^\A = t^a {A^a}^\A$ has been used.
In this expression, $y \equiv (y^+,y^-,y_\perp)$ and assuming a weak dependence on $Y^-$, 
we have replaced $Y^-$ with $0$ simply to save writing.  
Integrating over $y^+$ by parts, the overall factor of $k^-$ in $\mathcal{K}$ is 
converted into the derivative $- i  \prt^-$ acting on $\mathcal{Y}$. The terms in $\mathcal{K}$ are then ordered in 
powers of $q^-$, the forward energy of the jet, which represents the largest scale in the problem. 
We begin by analyzing the leading term, which yields the coefficient of the first 
derivative of the longitudinal momentum $\kd$-function as, 
\bea
C_1 &\propto& \int dY^- d^4 y \frac{d k^- d^2 k_\perp}{(2 \pi)^3} 
e^{-i \frac{k_\perp^2}{2 q^-} y^- - i k^- y^+ + i k_\perp \x y_\perp} \nn \\
 \ata  \frac{4 \pi \A}{ 2 N_c} \lc p_2 |  \left[ i \prt^- {A^a}^+(y) {A^a}^+ (0) \right] | p_2 \rc.
\eea
The `$\propto$' sign denotes that there is an over 
all factor that will arise from the decomposition of $C^{A}_{p,p_2}$ in Eq.~\eqref{W_mu_nu} [see Refs.~\cite{Majumder:2007hx,Majumder:2008jy} for details]. The coefficient ($C_1$) 
above may be further simplified by ignoring the small factor $k_\perp^2/(2q^-)$ in the exponent and integrating out 
the $k_\perp$ and the $k^-$ to yield $\kd$-functions over $y^+$ and $y_\perp$.

Unlike the case of transverse broadening~\cite{Majumder:2007hx}, the two gluon matrix element is not 
manifestly gauge invariant. It may be cast in a gauge covariant form by noting that the gauge field at 
$y^+ \ra \infty $ is vanishingly small. In this limit,
\bea
%
%\prt^- A ^+ &\simeq& F^{-+} \Rightarrow \\
%
{A^a}^{+} (Y^-) &=& \int_{-\infty}^{0^+} d z^+ \prt^- {A^a}^{+} (Y^-,z^+)  \label{F+-def}\\
&=& \int_{-\infty}^{0} d z^+ {F^a}^{-+} (Y^-,z^+). \nn
\eea
%
% where, the standard notation of $A^\A = t^a {A^a}^\A$ has been used. 
This yields a somewhat complicated expression for the elastic energy loss coefficient. In the limit of 
$q^- \ra \infty$, where one may ignore the $k_\perp^2/(2 q^-)$  in  the exponent, the coefficient is given as, 
\bea
C_1 &=& \int dY^- d y^- d y^+ 
%\frac{d k^- d^2 k_\perp}{(2 \pi)^3} 
%
% e^{-i \frac{k_\perp^2}{2 q^-} y^- - i k^- y^+ + i k_\perp \x y_\perp} \nn \\
 %
%  \ata 
\frac{4 \pi \A}{ 2 N_c }  \frac{\rho_N}{2 p^+}  \label{e_loss_nuclei} \\
\ata \lc p_2 |  \left[ i {F^a}^{- +} (y^-,0,0)  {F^a}^{- +} (0,y^+,0) \right] | p_2 \rc. \nn
\eea
%
% In the equation above, $F^{-+} = t^a {F^a}^{-+}$ and the trace is meant over color. 
The factor of $\rho_N/(2p^+)$, where, $\rho_N$ is the nucleon density and $p^+$ is the average forward 
momentum of the nucleon, is obtained from a factorization of the correlation coefficient 
$C_{p,p_2}^A \simeq C_p^A \times \rho_N/(2p^+)$, which is similar to the procedure used in Refs.~\cite{Majumder:2007hx} to 
indicate the weak correlation between nucleons.

In the derivation of the leading contribution to elastic energy loss in large nuclei, 
Eq.~\eqref{e_loss_nuclei} is as far as one may proceed without invoking a model of the 
gluon distribution within the nucleons. In the case of transverse broadening~\cite{Majumder:2007hx}, 
a similar distribution 
involving the gauge field $F^{+ \perp}$ results; this is combined with the nucleon density $\rho_N$ 
to define the transport coefficient $\hat{q}$. In transverse broadening and energy loss in both large 
nuclei and heavy-ion collisions, $\hat{q}$ is often used as a parameter to fit with experimental 
data~\cite{Majumder:2007ae}. We anticipate that a similar coefficient may also be motivated for the elastic energy loss, denoted as 
\bea
\hat{e} &=&  \frac{4 \pi \A}{ 2 N_c  }  \frac{\rho}{2 p^+}   \\
\ata \!\!\!\int dy^- dy^+  \lc p | \left[ i {F^a}^{- +} (y^-,0,0)  {F^a}^{- +} (0,y^+,0) \right] | p \rc. \nn \label{e_loss_nuclei_fact}
\eea 
The coefficient above represents the leading contribution to the elastic energy loss per unit length as encountered by 
a hard jet due to soft rescattering in the medium. There exist multiple other contributions such as 
energy loss due to hard rescattering, Compton scattering off hard gluons  and subleading contributions 
(in $q^-$ power counting) 
from Eq.~\eqref{k_expand}. The first two contributions were calculated in Ref.~\cite{Wang:2006qr} and the 
emission of a hard forward parton in the final state leads to an interference of such contributions with 
radiative energy loss. 
As a result, these contributions are appropriately combined with and should be considered as a part of the radiative energy 
loss calculation~\cite{HT}. 
In this manuscript,  the focus has been restricted only on those sub-leading contributions which do not have 
a hard on-shell gluon in the final state and thus do not interfere with radiative energy loss. Alternatively 
stated, we are only considering contributions where the parton loses longitudinal momentum $q^-$ 
without producing another hard parton with large ($-$)-component of light cone momentum.

As mentioned above, the effect of energy loss due to soft scattering off the glue field has been completely
factorized from the production process of the hard jet. As a result, the coefficient $\hat{e}$ may be 
computed in any medium. 
In the case of elastic energy loss in a high temperature plasma, the coefficient in the equation above 
may be calculated exactly. Following the methods outlined in Ref.~\cite{Majumder:2007hx}, we reabsorb the factor 
$ \rho/(2 p^+) $ into the definition of a generalized medium state $| n \rc$ 
and define the elastic loss coefficient as the expectation of the operator $i \prt^- A^+(y^-) A^+(0)$
in the ensemble of thermal states as, 
\bea
\mbx\!\!\!\!\!\!\!\!\!\!\hat{e}_{HTL} \!\!\!\!&=&\!\!\!\! \frac{ 4 \pi \A \int d y^- \lc n | e^{-\B \hat{H}} [ i \prt^- {A^a}^+(y^-) {A^a}^+(0) ] | n  \rc
}{ 2 N_c } .  \label{e_loss_thermal}
\eea
Using the standard decomposition of two-point operators in finite temperature field theory~\cite{Kapusta:2006pm}, 
we decompose the correlator in the 
equation above, in the HTL limit, as, 
\bea
\hat{e}_{HTL} &=& \frac{4 \pi \A_s }{2N_c} \int \frac{dy^-d^4 k}{(2 \pi)^4} 
e^{-i k^+ y^- } \nn \\ 
\ata  k^- \rho^{++}_{a b}
 [1 + n_B(k^0)] \kd^{ab} ,\label{e_spec_den}
\eea
where, $n_B$ is the Bose distribution function and 
$\rho_{ab}^{++} $ represents the (++)-component of the gluon spectral density which is diagonal in 
color space, ($\rho_{ab} = \rho \kd_{ab}$).

The spectral density has the usual decomposition in terms of 
transverse and longitudinal components (using covariant gauge):
\bea
\rho^{++} &=& P_T^{++ } \rho_T + P_L^{++ } \rho_L \label{spec_den_proj}
\eea
The projectors are further simplified on performing the $y^-$ integral which yields $\kd(k^+)$. As a result, 
$k^-  = - 2k^3$ and $P_T^{++} = P_T^{33} = - P_L^{33} = | k_\perp^2 |/ | (k^3)^2  + k_\perp^2|$. 
Using the standard notation ($k^0/|\vk|=x$)~\cite{Kapusta:2006pm}, the expression for the elastic energy loss per unit length is given as,
\bea
\mbx\!\!\!\!\hat{e}\!\!\!&=&\!\! 4 \pi \A_s\!\!\!\!\!\int\limits_0^{Q^2_{MAX}} \frac{d|\vk|^2 }{ 2 \pi  } \int\limits_0^1 \frac{dx}{2\pi} 
\frac{|\vk|^2 (-x) (1-x^2)(N_c^2 - 1) }{4 N_c}    \label{e_final}\\
\mbx\!\!\!\ata\!\!\!\! \left[ \left\{ \rho_T (|\vk|,k^0)  - \rho_L (|\vk|,k^0) \right\} \left\{ 1 + n_B(k^0) \right\} \right]_{k^0 = -|\vk|x} . \nn
\eea 

The equation above for the loss of light cone momentum $l^-$ per unit of light cone path $L^-$ is 
equivalent to the energy lost by near on-shell partons per unit length \ie., $\hat{e} \simeq dE/dL$. 
It may be evaluated numerically using the actual form of the gluon spectral density. In realistic 
calculations of the elastic energy loss at finite temperature, the form of the spectral density 
depends on the exchanged momentum of the gluon. For hard exchanges, one may simply evaluate the 
$t$-channel matrix elements for the scattering of a hard jet parton off a hard ($\sim T$) medium 
parton. These are logarithmically divergent and are regulated at a soft scale $\mu_T^2$. Soft 
exchanges below this scale are appropriately calculated within the HTL formalism. On combining 
both these computations the arbitrary scale $\mu_T^2$ is removed (see Refs.~\cite{Thomas:1991ea,Qin:2007rn} for 
details and results of this procedure).

\begin{figure}[htbp]
%\begin{center}
%  \epsfxsize 80mm
%\hspace{0cm}
\resizebox{3in}{2.5in}{\includegraphics[0in,0in][4.5in,4.5in]{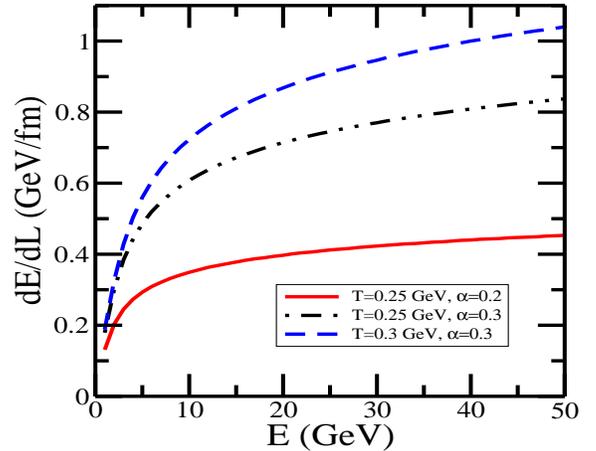}} 
%\vspace{0.25cm}
    \caption{(Color Online) Elastic energy loss per unit length ($dE/dL$) as a function of the energy of the 
propagating parton for different fixed temperatures of the media and values of $\A_s$. See text for 
details.} 
    \label{fig1}
%  \end{center}
\end{figure}

The goal of this Note is not to rederive the results of Refs.~\cite{Thomas:1991ea,Qin:2007rn} 
in their entirety but 
simply to show that a similar procedure may be carried out here with identical results. 
Adjusting for the choice of gauge and removing the color factor $C_F$ for the case of QED, 
Eq.~\eqref{e_spec_den} may be shown to be equivalent to Eq.~(38) of the last article in Ref.~\cite{Thomas:1991ea}.
For the case of 
soft momentum transfers, one may use the known form of the HTL spectral densities~\cite{Kapusta:2006pm} 
to compute $\hat{e}$. 
%It is a trivial matter to show that the result of this substitution is equivalent to 
%Eq.~(38) of the last article of Ref.~\cite{Thomas:1991ea} (after adjusting for the choice of gauge and 
%removing the factor of $C_F$ for the case of QED energy loss).   
Within this approximation, the upper limit of the $|\vk|$ integral is terminated 
at an appropriate scale of $Q_{MAX} \sim T$. With a choice of 
an $\alpha_s = 0.3$ and a Debye mass of $m_D \simeq 4\pi \alpha_S T$, this yields an $\hat{e} \sim 0.06$ GeV/fm 
at a $T=300$ MeV. In Refs.~\cite{Thomas:1991ea,Qin:2007rn}, the spectral density is rederived for the 
case of hard momentum transfers; the difference between the results obtained from such a procedure and that from 
arbitrarily extending the HTL form of the spectral density to large momentum transfers is small ($\sim 20$\%) 
at large jet energy E ($\sim 40 - 50$ GeV)~\cite{qin_phd}. 
In this effort, we present this simplified estimate of the total elastic energy loss in a thermalized QCD medium by extending 
$Q_{MAX} \sim \sqrt{ET}$. This yields an energy dependent elastic energy loss per unit length as shown in Fig.~\ref{fig1}. 
The plot includes three different choices of $T$, and the coupling $\A_s$. The largest temperature and 
coupling yields the largest energy loss per unit length ($\hat{e}$), with $\hat{e}$ dropping when the temperature 
or the coupling is reduced. 
%Comparing with Ref.~\cite{Qin:2007rn}, the results obtained are approximately 25\% larger 
%than what would be obtained from calculation where the case of hard momentum transfers is treated separately~\cite{qin_phd}.
%
Previous 
efforts using the higher twist formalism ignored this kind of mechanism of elastic energy loss. 
The current manuscript justifies the inclusion of such contributions within the basic formalism of higher twist 
elastic energy loss. Unlike the case of transverse broadening, there does not seem to be a simple 
classical  analogue of this calculation, however, attempts in this direction 
%from kinetic theory 
are 
currently underway~\cite{Neufeld:2008fi}.

A closely related calculation, at the same order of expansion in $\A_s$, is  the 
fluctuation of the elastic energy loss. This is obtained as the coefficient of the  second derivative 
of the delta function of Eq.~\eqref{delta_func}.  Following a procedure almost identical to the 
steps carried out above, we obtain this coefficient in a thermal medium as, 
\bea
\hat{e}_2 &=&  \frac{ 4 \pi \A \int d y^- \lc n | e^{-\B \hat{H}}  [ {F^a}^{-+} {F^a}^{- +} ] | n  \rc }{ 2 N_c } ,
\eea
where, ${F^a}^{-+}$ has been defined in Eq.~\eqref{F+-def}. Following a resummation procedure, as in Ref.~\cite{Majumder:2007hx}, and by assuming that the correlation between $F^{+ \perp} $  and $F^{+-}$ is 
vanishing, one may immediately  postulate the diffusion equation for elastic energy loss as 
\bea 
\frac{\prt \phi (L^- , l^-)  }{\prt L^-} &=&  \hat{e} \frac{\prt \phi (L^- , l^-)  }{\prt l^- } 
+ \hat{e}_2 \frac{\prt^2 \phi (L^- , l^-)  }{\prt {(l^-)}^2 } , \label{diff_drag}
\eea
where the diffusion and drag coefficients have been evaluated above.

In the interest of completeness, we will also evaluate the next-to-leading terms in Eq.~\eqref{k_expand}
which provide additional contributions to the elastic energy loss. These are however, 
further suppressed by the hard scale of $q^-$. 
The next-to-leading contribution to the elastic energy loss or expectation of $k^-$ emanates from the contraction of the second term in $\mathcal{K}$ with $\mathcal{Y}$ from Eq.~\eqref{k_expand}. Evaluating this contribution, we obtain the 
coefficient of the first derivative of the $l^-$ distribution as
\bea
C_1 &\propto& \!\!\frac{-g^2}{2q^-} \int dY^- dy^-  \lc p | \prt^-\!A^+ (y^-/2) \prt^-\!A^+ (-y^-/2) | p \rc  \nn \\
%
%&+& g_\perp^{\A \B} \lc p | \prt^-\!A_\A (y^-/2) \prt^+\!A_\B (-y^-/2) \nn \\ 
%
%&+& 
%
% \prt^+\!A_\A (y^-/2) \prt^-\!A_\B (-y^-/2) | p \rc \nn \\
%
&\simeq&  \!\!\frac{-g^2}{2q^-} \!\!\int\!\! d Y^-\!\!dy^-  \lc p |  F^{- +} (y^-/2) F^{- +} (-y^-/2) | p \rc,
\eea
where, the standard shorthands of $A^\A = t^a {A^a}^\A$ and $F^{-+} = t^a {F^a}^{-+}$ have been used and a trace 
over color is implied. This term actually results in a slight gain in the longitudinal momentum fraction. 
In the derivation above, we have approximated that in the Breit frame 
(and in a covariant gauge) $A^+ \gg A_\perp \gg A^-$~\cite{lqs}  
and $\prt^- A^+ \simeq F^{- +} $.
While these contributions are suppressed by the large energy of the 
jet $q^-$, the operator products are not very different from those which constitute the leading 
contributions to elastic energy loss. As a result, such terms are only important at lower jet 
momenta.

In this Note, we have presented an extension of the higher-twist expansion formalism of jet modification 
to include the effect of elastic energy loss. This was carried out by extending the formalism of transverse 
broadening in Ref.~\cite{Majumder:2007hx}, by generalizing the two dimensional distribution of the 
propagating quark's transverse momentum to a three dimensional distribution including also its 
longitudinal momentum $l^-$. The equation governing the distribution in $l^-$ includes both a 
diffusion term and a drag term. In contrast to other formalisms of jet modification, both these 
terms were evaluated at the operator level, independent of the details of the medium. The sole 
assumption used was that the color correlation length in the medium is small. The drag 
coefficient which yields the elastic energy loss per unit length was then evaluated in a thermal 
plasma in the HTL limit. The results obtained are consistent with similar calculations in 
other formalisms~\cite{Thomas:1991ea,Qin:2007rn,Djordjevic:2006tw}. 
The final results, however, differed from previous attempts within the higher-twist formalism such as 
Ref.~\cite{Wang:2006qr} where elastic energy loss amplitudes  interfered with radiative 
processes. In this Note, the focus has been solely restricted to those 
processes which do not interfere with radiative amplitudes. 

The author thanks U.~Heinz, B.~M\"{u}ller  and G-Y.~Qin for extensive discussions. This work was supported in part by the U.~S.~Department of Energy, under grant nos. DE-FG02-05ER41367 and  DE-FG02-01ER41190.

\end{document}